**Kolmogorov-Arnold Networks Applied to Materials Property Prediction**

**Authors:** Ryan Jacobs[1], Lane E. Schultz[1], Dane Morgan[1]

[1] Department of Materials Science and Engineering, University of Wisconsin-Madison, Madison, WI, USA.

**Abstract**

Kolmogorov-Arnold Networks (KANs) were proposed as an alternative to traditional neural network architectures based on multilayer perceptrons (MLP-NNs). The potential advantages of KANs over MLP-NNs, including significantly enhanced parameter efficiency and increased interpretability, make them a promising new regression model in supervised machine learning problems. We apply KANs to prediction of materials properties, focusing on a diverse set of 33 properties consisting of both experimental and calculated data. We compare the KAN results to random forest, a method that generally gives excellent performance on a wide range of properties predictions with very little optimization. When using the standard 2-layer KAN architecture motivated by the Kolmogorov-Arnold theorem, we found, on average, KANs performed worse than random forest. The KANs were worse, on par, or better than random forest about 35%, 60%, and 5% of the time, respectively, and KANs are in practice more difficult to fit than random forest. By tuning the network architecture, we found property fits often resulted in 10-20% lower errors compared to the standard KAN, and typically gave results comparable to random forest. In the specific context of predicting reactor pressure vessel transition temperature shifts, we explored the parameter efficiency and the interpretable power of KANs by comparing predictions of simple KAN models (e.g., < 50 parameters) and closed-form expressions suggested by the KAN fits to previously published deep MLP-NNs and hand-tuned models created using domain expertise of embrittlement physics. We found that simple KAN models and the resulting closed-form expressions produce prediction errors on par with established hand-tuned models with a comparable number of parameters, and required essentially no domain expertise to produce. Overall, these findings reinforce the broad potential applicability of KANs for machine learning in materials science and suggest that KANs should be explored as a standard regression model for prediction of materials properties.

## 1. Introduction:

Multi-layer perceptrons (MLPs) have been the standard approach for constructing neural network-based machine learning (ML) models for decades and form the basis of all current deep learning-based neural network models. Recent work from Liu et al.[1] revitalized research on the Kolmogorov-Arnold Networks (KANs), expanding the original Kolmogorov-Arnold approach to be more general, flexible and conducive to how users perform ML research today. MLPs and KANs both consist of a feed-forward architecture with an input layer, one or more hidden layers, and an output layer. The central difference between MLPs and KANs is how the representation between inputs (features) and output (target) is learned. In MLPs, linear functions consisting of weights and biases for each node are learned. An activation function, which is typically the same for all nodes within a layer, is then applied to the output of these linear combinations. However, in KANs, it is effectively the activation functions themselves that are the target of the learning, with the form represented by a flexible parameterization, e.g., with spline models. The work of Liu et al. demonstrated impressive performance of KANs vs. MLPs for numerous synthetic numerical datasets as well as a select physics-based examples. One striking result from Liu et al. was the observed improvement in accuracy and parameter efficiency for solving a partial differential equation using KANs vs. using MLPs: they found the KAN model to be 100× more accurate than the MLP while simultaneously having 100× fewer parameters. KANs can be arbitrarily complex but we here define a "standard KAN" as consisting of 2 layers, with layer 1 having N nodes and layer 2 having 2N+1 nodes, for an input with N features. We call this standard as it is the simplest structure guaranteed to be able to represent any function of N variables by the Kolmogorov-Arnold theorem. However, as pointed out by Liu et al.[1], there is no assurance that this minimal standard KAN is optimal for achieving a robust efficient fitting, so other KAN architectures are generally important to explore.

The work of Liu et al. has prompted a frenzy of recent advancements and applications of KANs. For example, KANs have been adapted to handle temporal data for time-series forecasting,[2–5] modified to enable the use of graph-based KANs,[6–10] investigated as a potential image classification and segmentation model for computer vision tasks,[11–13] and more. [12]A recent reimplementation of KANs, called FastKAN, modifies the spline fitting to use radial basis functions, resulting in faster training than the initial KAN implementation with essentially no loss in accuracy.[14] The use of such basis functions and the graph-based implementation of KANs may open the door for these models to be used in many materials applications, e.g., as machine learning interatomic potentials.[15,16] In addition, KANs have been benchmarked against MLP NNs for a number of tabular datasets in Ref. [17] In that work, it was found that KANs perform on par or better than MLPs for a given number of model parameters, but at the expense of greater compute time for a KAN when compared to a MLP NN with the same number of parameters.[17]

Given the recent development of KANs, not much is known regarding how they perform in the domain of materials science, and in particular for the important task of materials property prediction. Despite this, there have been a handful of recent studies comparing the performance of MLPs vs. KANs on a few different materials properties, which we discuss below. Constructing ML models to predict materials properties is a key pillar of research in materials science, and useful for accelerating computational discovery and design of new materials prior to engaging in more expensive and time-consuming simulations and experiments.[18] Recently, Bandyopadhyay et al. applied KANs to the prediction of high entropy alloy properties.[19] As a first example, they compared the performance of KAN vs. MLP for classification of a high entropy carbide composition being single phase, where they find their KAN has perfect performance (F1=1.0). They find that an MLP with a similar number of parameters has excellent but slightly lower performance, with F1=0.97. As a second example, they compare the performance of KAN and MLP for predicting yield strength and ultimate tensile strength. For models with similar numbers of parameters, KAN and MLP achieve ultimate tensile strength test RMSE values of 126 and 130 MPa, respectively, and yield strength test RMSE values of 136 and 155 MPa, respectively. While MLP models with larger numbers of parameters were found to outperform their KAN models, Bandyopadhyay et al. effectively showed that KANs and MLPs are competitive, while KANs are more readily interpretable than MLPs. In a third study, Fronzi et al.[20] compared the performance of KANs and MLPs for predicting band gaps and Seebeck coefficients for the design of thermoelectric materials. Like Bandyopadhyay et al., Fronzi et al. also find comparable performance of KANs and MLPs, where they obtained KAN and MLP test RMSEs of 0.146 eV and 0.150 eV for band gap, respectively, and KAN and MLP test RMSEs of 69.32 μV/K and 73.62 μV/K for Seebeck coefficient, respectively. As a final example of comparing KAN and MLP performance, Pourkamali-Anaraki[21] analyzed the breast cancer dataset from scikit-learn, and found that the MLP model had slightly higher median classification accuracy (about 97%) compared to the KAN model (about 95%), despite the KAN model having an order of magnitude more learnable parameters. When analyzing MLP vs. KAN performance on a small dataset (only 104 samples) of a dataset of 3D printer type prediction, they found that MLPs had median accuracies of about 90%, while the KANs struggled with median accuracies of only 40-53%, depending on the number of hidden layers used. Interestingly, there was much more variability in the KAN accuracy over the various cross validation folds, while the MLP model showed more consistent performance. Overall, these studies found that MLPs and KANs are generally competitive regarding typical test data errors and KANs are more interpretable than MLPs. However, KANs often show greater performance variation over a random set of cross validation splits and may in practice be more difficult to fit due to increased sensitivity to the exact architecture and hyperparameter choice compared to MLPs.

There are two major motivations for exploring KANs in materials. The first motivation is the possibility that they are numerically superior to standard methods (e.g., random forest or MLPs) in some some basic metric, where the most important would be required training data for a given accuracy or accuracy for a given set of training data. Effectiveness on these metrics is particularly important to explore since KANs have shown an exceptional ability to efficiently ferret out relatively simple functional relationships of the types often found in physical systems. Another motivation is that KANs can produce highly interpretable outputs and such interpretability is often quite valuable in scientific contexts. Finally, we note that KANs often seem to be able to use fewer parameters than other models, which is not in itself a very significant advantage, but tends to correlate with less overfitting and more accurate prediction (particularly for extrapolations on data further from the training data), greater interpretability, smaller training data requirements, and faster speed of inference. It is worth noting that recent studies have focused on comparing the performance of KANs with MLPs. However, in materials property prediction, small datasets with order a few hundred to a thousand data points are often better fit, and almost always more easily fit, using a model like random forest as opposed to an MLP, indicating that comparing KANs with random forest would be useful. With these overall motivations in mind, we here target three key questions for the application of KANs in the context of materials property prediction: (1) How do standard KANs with no network architecture optimization, that is, those using the standard 2-layer architecture suggested by the Kolmogorov-Arnold theorem, compare with other methods that can be applied essentially out-of-the-box with little to no hyperparameter tuning (e.g., random forest models)? (2) How do KANs with modest optimization of hyperparameters and network architecture perform relative to the default KAN architecture, and relative to random forest models? (3) Can we use KANs to reduce complexity and aid in interpretability of materials property prediction?

This work explores a number of data sets to help answer the above three questions. Regarding (1), we find that standard KANs with no optimization and employing 2-layer architectures in accordance with the Kolmogorov-Arnold theorem are generally competitive with, but on average worse than, random forest models for materials property prediction. Regarding (2), we find that optimization of hyperparameters and KAN network architecture results in modest improvement of fits, on average of about 10% lower errors than non-optimized KANs, and that on average the optimized KAN and random forest accuracies are about the same. Regarding (3), we show on a single materials property dataset of reactor pressure vessel steel embrittlement the advantages of KANs relative to deep MLP NN models. First, we find KANs to have competitive errors with deep MLP NNs while using 100× fewer parameters, a finding which is consistent with the understanding that KANs tend to be more parameter efficient than MLPs.[1] Second, we show that very small KAN models, containing fewer than 50 parameters, are capable of rivaling hand-tuned analytical models in terms of prediction accuracy. With these small KAN models, the ability to fit closed-form formulas to the KAN activation functions makes the model more readily interpretable than MLPs.

# 2. Data and Methods:

## 2.1. Data:

The datasets used in this work were collected and analyzed from a previous study.[22] The datasets consist of a total of 33 different materials properties, contain both experimental and calculated data, and the data types, dataset sizes, types and numbers of features, and original data references are summarized in **Table 1**. Additional details of the feature generation and selection procedure are provided in Ref. [22], and the datasets used here and in Ref. [22] are available online using the data links provided in Ref. [22]. We note that all datasets used in this work are available as part of Ref. [22] and are also hosted for easy access on Foundry-ML.[23] Note that none of these datasets use atomic structure in its featurization. The effectiveness of KANs (and their generalizations, e.g., graph-based KANs[6–9]) for this type of featurization is an important area of active and future study, but not the focus of this work.

**Table 1.** Summary of materials property datasets investigated in this work. Property names are listed in alphabetical order. Abbreviations: $R_c$ = critical cooling rate, $D_{max}$ = maximum casting diameter, ASR = area specific resistance. TEC = thermal expansion coefficient, RPV = reactor pressure vessel, $T_c$ = superconducting critical temperature.

| Property name | Data type | Number of data points | Feature type | Number of features | Data reference |
|---|---|---|---|---|---|
| Bandgap | Experiment | 6031 | Elemental | 25 | [24] |
| Concrete compressive strength | Experiment | 1030 | Material-specific | 8 | [25] |
| Debye Temperature | Computed | 4896 | Elemental | 25 | [26] |
| Dielectric constant | Computed | 1056 | Elemental | 25 | [27] |
| Dilute solute diffusion | Computed | 408 | Elemental | 25 | [28] |
| Double perovskite bandgap | Computed | 1306 | Elemental | 25 | [29] |
| Elastic tensor (bulk modulus) | Computed | 1181 | Elemental | 25 | [30] |
| Exfoliation energy | Computed | 636 | Elemental | 25 | [31] |
| Heusler magnetization | Computed | 370 | Elemental + one-hot | 25 | [32] |
| | | | encoding of Heusler structure type | | |
| High entropy alloy hardness | Experiment | 370 | Elemental | 25 | [33] |
| Lithium conductivity | Experiment | 372 | Elemental | 25 | [34] |
| Metallic glass $D_{max}$ | Experiment | 998 | Elemental | 25 | [35] |
| Metallic glass $R_c$ (from LLM) | Experiment | 297 | Elemental | 25 | [36] |
| Metallic glass $R_c$ | Experiment | 2125 | Elemental + one-hot encoding of data type | 25 | [37] |
| Mg alloy yield strength | Experiment | 365 | Material-specific | 14 | [38] |
| Oxide vacancy formation | Computed | 4914 | Elemental | 25 | [39] |
| Perovskite ASR | Experiment | 289 | Elemental + one-hot encoding of electrolyte type + ML-predicted barrier | 25 | [40] |
| Perovskite conductivity | Experiment | 7230 | Elemental + material-specific | 25 | [41] |
| Perovskite formation energy | Computed | 9646 | Elemental | 25 | [42] |
| Perovskite H absorption | Experiment | 795 | Elemental + material-specific | 25 | [43] |
| Perovskite O p-band center | Computed | 2912 | Elemental | 50 | [44] |
| Perovskite stability | Computed | 2912 | Elemental | 50 | [44] |
| Perovskite TEC | Experimental | 137 | Elemental | 25 | [45] |
| Perovskite work function | Computed | 613 | Elemental | 25 | [46] |
| Phonon frequency | Computed | 1265 | Elemental | 25 | [47] |
| Piezoelectric max displacement | Computed | 941 | Elemental | 25 | [48] |
| RPV transition temperature shift | Experiment | 4535 | Material-specific + one-hot encoding of product form | 15 | [49] |
| Semiconductor defect levels | Computed | 896 | Elemental + material-specific | 25 | [50] |
| Steel yield strength | Experiment | 312 | Elemental | 25 | [51] |
| Superconductivity $T_c$ | Experiment | 6252 | Elemental | 50 | [52] |
| Thermal conductivity | Experiment | 872 | Elemental + material-specific | 25 | n/a |
| Thermal conductivity (AFLOW) | Computed | 4887 | Elemental | 25 | [26] |
| Thermal expansion | Computed | 4886 | Elemental | 25 | [26] |

## 2.2. Model fitting

All KAN models were fit using the *pykan* python package (version 0.0.5), which is the official KAN implementation from the work of Liu et al.[1] During the course of this work, the pykan package was updated numerous times. As of this writing, the latest version is 0.2.8. We performed fits using v 0.0.5 and 0.2.8 to the dilute solute diffusion and the concrete compressive strength datasets, and found no statistically significant change to the fit results. We also integrated the KAN model into our MAterials Simulation Toolkit for Machine Learning (MAST-ML) software,[22,53] enabling users to fit KAN models and also draw upon the numerous materials property-centric analyses and automation present in MAST-ML (e.g., featurization methods, plotting, uncertainty quantification, etc.)

The fits discussed in **Section 3.1** use non-optimized, 2-layer standard KAN models. [1]For example, a dataset with 25 elemental property features will have an architecture of 25-51-1, meaning there are 25 input nodes, a single hidden layer with 51 nodes, and a single output node. The performance of KANs is evaluated by performing 25 splits of leave-out 20% cross validation, and the average and standard deviation of the root-mean-squared-error (RMSE) and reduced RMSE (RMSE/$\sigma_Y$, where $\sigma_Y$ is the standard deviation of the entire dataset) are tabulated. We note that data on all fits and associated metrics ($R^2$, mean absolute error, RMSE, RMSE/$\sigma_Y$) are available as part of the digital supporting information. All fits use normalized input of features, where the normalization is done with the StandardScaler tool in scikit-learn.[54] These values are then compared with the same statistics for a random forest model constructed using the same feature set and 25 splits of 5-fold cross validation. We noticed that for some splits, KAN will have some numerical instability which leads to very high test RMSEs. To circumvent this issue, we examine the list of test RMSEs and remove those which have values greater or less than two standard deviations from the mean test RMSE, where this standard deviation of the test RMSE is calculated using all of the data. For each dataset, two runs were performed: one where the KAN was trained for 10 iterations, and another where the KAN was trained for 20 iterations. We initially started with 20 iterations, but the runs with 10 iterations were added because we noticed that results for some properties were showing significant overfitting.

In **Section 3.2**, we perform modest manual tuning of KAN architecture. For these runs, only the network architecture (number of nodes per layer, number of hidden layers) was changed, while the same number of grid intervals (grid=5) and same order of the spline piecewise polynomial degree (k=3) was used for all cases. For this manual tuning, we first constructed 2-layer KANs with node counts varying from 1 up to 2N+1. The best performing 2-layer KAN architecture was then extended to 3 layers, with the node count in the third layer gridded from 1 to 110. These parameter ranges are somewhat arbitrary, and were used as a means to assess whether the standard 2-layer KAN may be improved with modest effort. Additional tuning of KAN models using more automated approaches is discussed in **Section 3.3**.

In **Section 3.3**, we perform automated hyperparameter searches using the Optuna[55] python package and compare these automated optimized network fits to our initial standard KAN model and fits to random forest models in the literature. For a given set of hyperparameters, a KAN model was trained on a random split of 80% of the data and then validated on the remaining 20%, and this process was performed five times. The lowest RMSE from these five splits was used to evaluate the performance of the set of hyperparameters used. This process was repeated for 100 sets of hyperparameters, and the optimal set of parameters with the lowest RMSE was used for the final optimized model. The hyperparameters optimized included the grid interval (ranging from 5 to 500, this controls the density of points for each spline function), regularization weight (ranging from 1e-10 to 1e-3 in a logarithmic scale, this impacts the loss function by penalizing overly complex spline functions), the spline piecewise polynomial degree (ranging from 1 to 3, this controls the functional form of the spline functions), the number of hidden layers (ranging from 0 to 3), and the number of nodes per layer. We performed optimization on 24 datasets (9 were excluded because of large computational expense) by allowing the number of nodes to vary from 1 to 10 in each hidden layer. In addition, for a subset of 12 datasets, the number of nodes in each hidden layer was varied from 1 to 100. For each optimization run, a total of 100 parameter configurations were tested and each model was trained for 25 iterations. While these tests are by no means exhaustive, they were sufficient to enable us to gauge the typical change in prediction error resulting from at least modest parameter tuning relative to standard KAN fits. We acknowledge that our optimization approach may be prone to overfitting as the optimization should ideally be peformed on data not used in assessing the model, e.g., using nested cross-validation. However, a nested approach was not employed due to its prohibitive computational cost. To address this issue we checked for overfitting using the diffusion dataset, where we held out a 20% test set at the outset, then performed the manual parameter tuning by optimizing on cross-validation of the remaining 80% of the data. We found that the optimum network architecture changed slightly from that reported in **Section 3.2**, to 25-37-47-1, with a mean validation RMSE of 0.31 eV and left out dataset RMSE of 0.32 eV. This test shows that no significant overfitting occurred for the diffusion dataset and we therefore assume that overfitting during network optimization is not a big concern.

# 3. Results and Discussion:

## 3.1. 2-layer Standard KAN performance

In this section, we compare the performance of random forest models fit in previous works with 2-layer standard KAN models (see **Section 2.2** for exact definition of a "standard" KAN). **Figure 1** provides a summary of ML model fits with both 2-layer standard KANs and random forest models across all datasets examined in this work, while **Table 2** provides a full list of the RMSE and RMSE/$\sigma_y$ values for random forest and KAN models trained using both 10 and 20 training iterations (see **Section 2** for more details). Note, for **Figure 1**, we plot the KAN result with lowest RMSE/$\sigma_y$ value between the 10- and 20-iteration fits. Overall, we find that the performance of random forest vs. KAN varies based on the property under study; there are cases where random forest provides the best fit and others where KANs appear more accurate. While 33 datasets were explored in this work, two of them, superconductivity and thermal expansion (from AFLOW database) could not be adequately fit with KANs and are not discussed further here. Of the 31 datasets for which direct comparisons can be made, we find that 19 properties showed better fits with random forest, while 12 showed better fits with KANs. It is worth noting that when random forest resulted in the better model, its improvement in RMSE/$\sigma_y$ was more substantial than for the cases where KANs had the lower error. More specifically, for the 19 datasets with lower errors from random forest, the RMSE/$\sigma_y$ improvement vs. KANs was on average 36.3% in RMSE/$\sigma_y$. For the 12 datasets which showed lower errors with KANs, the improvement vs. random forest was, on average, 8.5% in RMSE/$\sigma_y$. Across all properties, the average (+/- cross validation standard deviation) random forest and KAN RMSE/$\sigma_y$ values were 0.418 +/- 0.062 and 0.485 +/- 0.029, respectively. We are aware of only one of these data sets, the concrete compressive strength, which has been fit using KANs by another research group. This work, from Wang et al.,[56] examined the performance of KANs for this concrete compressive strength dataset and obtained a test RMSE of 5.09 MPa. This test RMSE is quite close to our value of 5.21 MPa in **Table 2**. Overall, many properties showed fits that were very comparable between KANs and random forest, where 14 of the 31 properties had average RMSE/$\sigma_y$ values within 0.062 of each other (the average random forest cross validation error bar displayed in **Figure 1**). These findings demonstrate that KANs, in their standard 2-layer construction following the Kolmogorov-Arnold theorem, are valuable models for materials property prediction, even without further optimization. We now turn to exploring optimization of KANs in **Section 3.2** and **Section 3.3**.

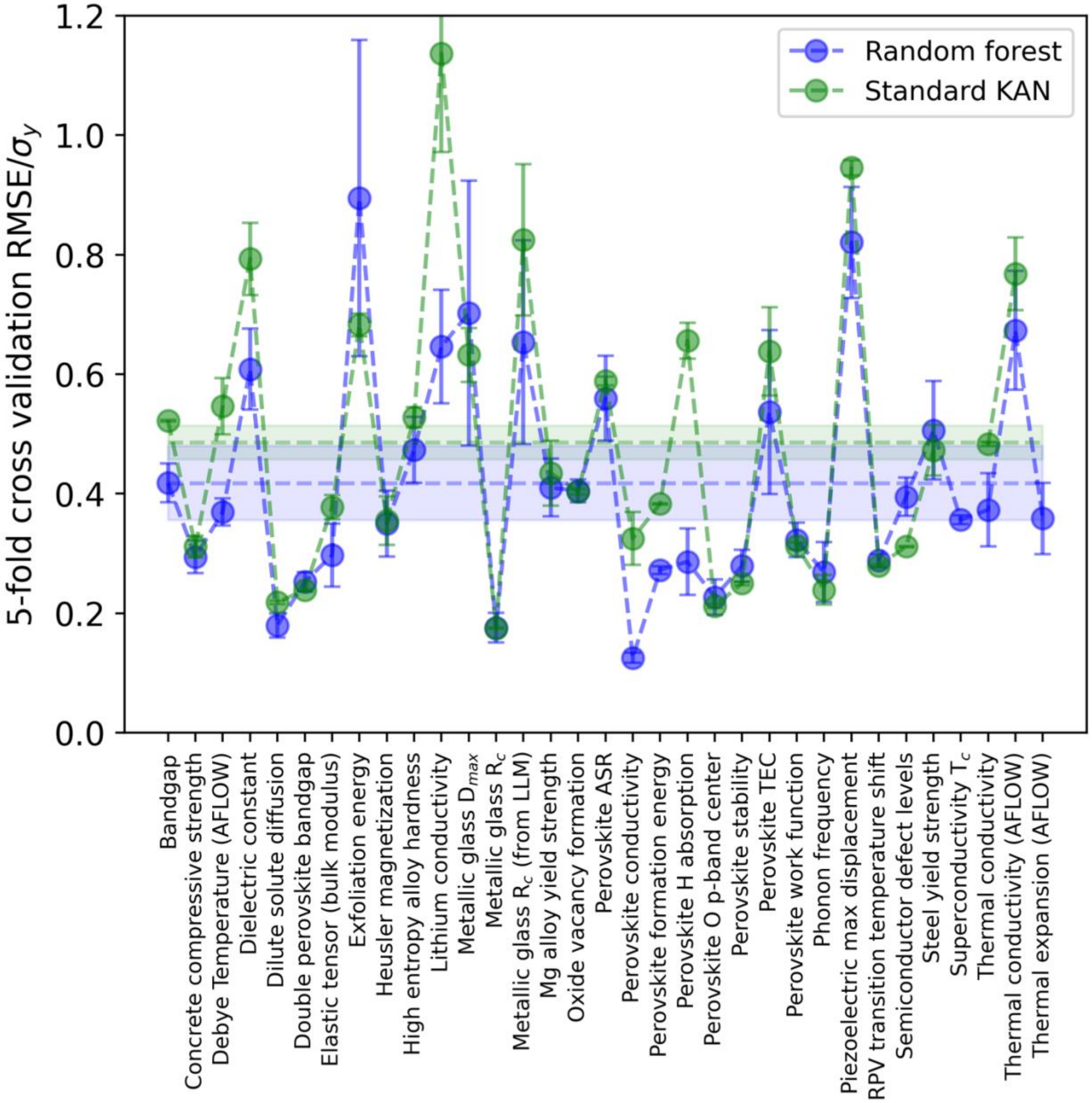

**Figure 1.** Summary of test data RMSE/$\sigma_y$ values for 2-layer standard KAN models (green data) and previously fit random forest models (blue data) for all datasets. The KAN results with lowest RMSE/$\sigma_y$ value between the 10- and 20-iteration fits are plotted here. The points and error bars represent mean and standard deviations, respectively, across all test data splits. The horizontal dashed green and blue lines are the average RMSE/$\sigma_y$ values for standard KAN and RF models, respectively, while the green and blue shaded regions are the average test data error bars from standard KAN and RF models, respectively, where the error bar for a single property is the standard deviation of RMSE/$\sigma_y$ values for all test data splits. Note that the superconductivity $T_c$ and AFLOW thermal expansion datasets could not be adequately fit with standard KANs and are excluded from this plot.

**Table 2.** Summary of materials property fits from **Figure 1** comparing random forest and 2-layer standard KAN models. The numbers are RMSE values, and the numbers in parentheses are RMSE/$\sigma_y$ values. Abbreviations: $R_c$ = critical cooling rate, $D_{max}$ = maximum cast diameter, ASR = area specific resistance. TEC = thermal expansion coefficient, RPV = reactor pressure vessel, $T_c$ = superconducting critical temperature.

| Property (units) | Random forest RMSE (RMSE/$\sigma_y$) | 2-layer KAN RMSE (RMSE/$\sigma_y$) (10 iterations) | 2-layer KAN RMSE (RMSE/$\sigma_y$) (20 iterations) |
|---|---|---|---|
| Bandgap (eV) | 0.65 (0.42) | 0.81 (0.52) | 1.11 (0.71) |
| Concrete compressive strength (MPa) | 4.91 (0.29) | 6 (0.36) | 5.21 (0.31) |
| Debye Temperature (K) | 68.54 (0.37) | 1206.09 (6.51) * | 101.31 (0.55) |
| Dielectric constant (log scale) (n/a) | 0.17 (0.61) | 0.24 (0.83) | 0.23 (0.79) |
| Dilute solute diffusion (eV) | 0.24 (0.18) | 0.36 (0.27) | 0.3 (0.22) |
| Double perovskite bandgap (eV) | 0.4 (0.25) | 0.41 (0.26) | 0.38 (0.24) |
| Elastic tensor (bulk modulus) (GPa) | 21.66 (0.3) | 27.38 (0.38) | 31.18 (0.43) |
| Exfoliation energy (eV/atom) | 120.04 (0.89) | 91.61 (0.68) | 135.13 (1.01) |
| Heusler magnetization (emu/cm$^3$) | 154.39 (0.47) | 171.48 (0.53) | 179.48 (0.55) |
| High entropy alloy hardness (HV) | 79.7 (0.35) | 80.87 (0.36) | 107.54 (0.47) |
| Lithium conductivity (log scale) (S/cm) | 1.35 (0.65) | 2.37 (1.14) | 25.21 (12.07) * |
| Metallic glass $D_{max}$ (mm) | 5.39 (0.70) | 4.86 (0.63) | 8.18 (1.06) |
| Metallic glass $R_c$ (log scale) (K/s) | 0.39 (0.18) | 0.39 (0.18) | 0.54 (0.24) |
| Metallic glass $R_c$ (from LLM) (log scale) (K/s) | 1.3 (0.65) | 88.08 (44.29) * | 1.64 (0.82) |
| Mg alloy yield strength (MPa) | 44.59 (0.41) | 47.23 (0.43) | 53.18 (0.49) |
| Oxide vacancy formation (eV) | 1.74 (0.4) | 1.73 (0.4) | 1.74 (0.41) |
| Perovskite ASR (log scale) (Ohm-cm$^2$) | 0.61 (0.56) | 0.69 (0.64) | 0.64 (0.59) |
| Perovskite conductivity (log scale) (S/cm) | 0.26 (0.13) | 0.67 (0.32) | 28984.92 (13773.8) * |
| Perovskite formation energy (eV/atom) | 0.15 (0.27) | 0.22 (0.41) | 0.21 (0.38) |
| Perovskite H absorption (mol/formula unit) | 0.02 (0.29) | 0.04 (0.66) | 0.31 (4.60) |
| Perovskite O p-band center (eV) | 0.27 (0.23) | 0.25 (0.21) | 0.29 (0.25) |
| Perovskite stability (meV/atom) | 53.55 (0.28) | 47.9 (0.25) | 59.63 (0.31) |
| Perovskite TEC ($\times 10^{-6}$ K$^{-1}$) | 2.21 (0.54) | 9.32 (2.3) | 2.63 (0.64) |
| Perovskite work function (eV) | 0.56 (0.32) | 0.59 (0.34) | 0.54 (0.31) |
| Phonon frequency ($cm^{-1}$) | 132.6 (0.27) | 118.87 (0.24) | 118.09 (0.24) |
| Piezoelectric max displacement (log scale) ($C/m^2$) | 0.62 (0.82) | 0.71 (0.95) | 0.73 (0.97) |
| RPV transition temperature shift (°C) | 13.29 (0.29) | 12.9 (0.28) | 13.15 (0.29) |
| Semiconductor defect levels (eV) | 0.51 (0.39) | 0.4 (0.31) | 0.5 (0.39) |
| Steel yield strength (MPa) | 152.52 (0.51) | 142.55 (0.47) | 198.27 (0.66) |
| Superconductivity $T_c$ (natural log scale) (K) | 0.26 (0.36) | 14.22 (19.65) * | 4 (5.52) * |
| Thermal conductivity (log scale) (W/m-K) | 0.2 (0.37) | 0.26 (0.50) | 0.25 (0.48) |
| Thermal conductivity (AFLOW) (W/m-K) | 6.98 (0.67) | 7.96 (0.77) | 9.82 (0.95) |
| Thermal expansion ($K^{-1}$) | $1.41\times10^{-5}$ (0.36) | 0.03 (773.46) * | 0.03 (734.89) * |

*Numerical issues led to poor model fits

### 3.2. KAN performance by manually tuning network parameters

The above results in **Section 3.1** focused on construction of standard KAN models based on the Kolmogorov-Arnold theorem. Here, we explore how modest manual tuning of the KAN architecture can alter the prediction results. For this analysis, we focus on a subset of seven materials properties: dilute solute diffusion, Heusler magnetization, high entropy alloy hardness, metallic glass $R_c$ values, perovskite stability, RPV transition temperature shifts, and steel yield strengths. These datasets were chosen because the 2-layer standard KAN and random forest fits are competitive, and it is of interest to ascertain whether tuning the KAN model may result in improved fits. A summary of test data RMSE values for fits to these datasets is presented in **Table 3**. **Table 3** compares the RMSE from 2-layer standard KANs and the best manually tuned KAN from our analysis of gridding 2-layer and 3-layer architectures. In **Figure 2**, we present a graphical comparison of the test data RMSE values between random forest, 2-layer standard KAN fits from **Section 3.1**, and the manually tuned KANs discussed in this section. In **Figure 2**, the RMSE values are normalized to be percent changes relative to the random forest errors. Interestingly, we find that KAN models constructed based on the standard Kolmogorov-Arnold theorem resulted in the best fit for none of the properties considered here. In addition, we found that adding a third KAN layer resulted in an improved fit compared to using two layers for five of the seven properties. We find that modestly tuning KANs resulted in an RMSE reduction ranging from 0.5% up to about 22%, with an average RMSE reduction of nearly 11% across the seven properties analyzed here. After tuning the KAN layers and node counts, four of these seven properties show notably lower RMSE values than the previously fit random forest model (high entropy alloy hardness, perovskite stability, RPV transition temperature shift, and steel yield strengths), one property is essentially identical with random forest (metallic glass $R_c$), and one property is still higher than random forest but was significantly improved by the layer tuning (dilute solute diffusion).

**Table 3.** Summary of 5-fold cross validation RMSE (RMSE/$\sigma_y$) results for manually tuned KAN fits to select materials properties. For direct comparison, all models were fit using 10 training iterations.

| **Property (units)** | **2-layer standard KAN fit (Kolmogorov-Arnold theorem)** | **Best manually tuned KAN fit** | **Best KAN architecture** | **Percent improvement vs. standard KAN model** |
|---|---|---|---|---|
| Dilute solute diffusion (eV) | 0.36 (0.27) | 0.28 (0.21) | 25-47-103-1 | 22.2 |
| Heusler magnetization (emu/cm$^3$) | 172.09 (0.53) | 161.27 (0.49) | 25-13-1 | 6.3 |
| High entropy alloy hardness (HV) | 80.9 (0.36) | 67.8 (0.30) | 25-15-101-1 | 16.2 |
| Metallic glass $R_c$ (log scale) (K/s) | 0.39 (0.18) | 0.39 (0.18) | 25-51-81-1 | 0.5 |
| Perovskite stability (meV/atom) | 47.9 (0.25) | 42.8 (0.22) | 50-85-71-1 | 10.6 |
| RPV transition temperature shift (°C) | 12.9 (0.28) | 12.4 (0.27) | 15-23-75-1 | 3.9 |
| Steel yield strength (MPa) | 142.55 (0.47) | 125.69 (0.42) | 25-19-1 | 11.8 |

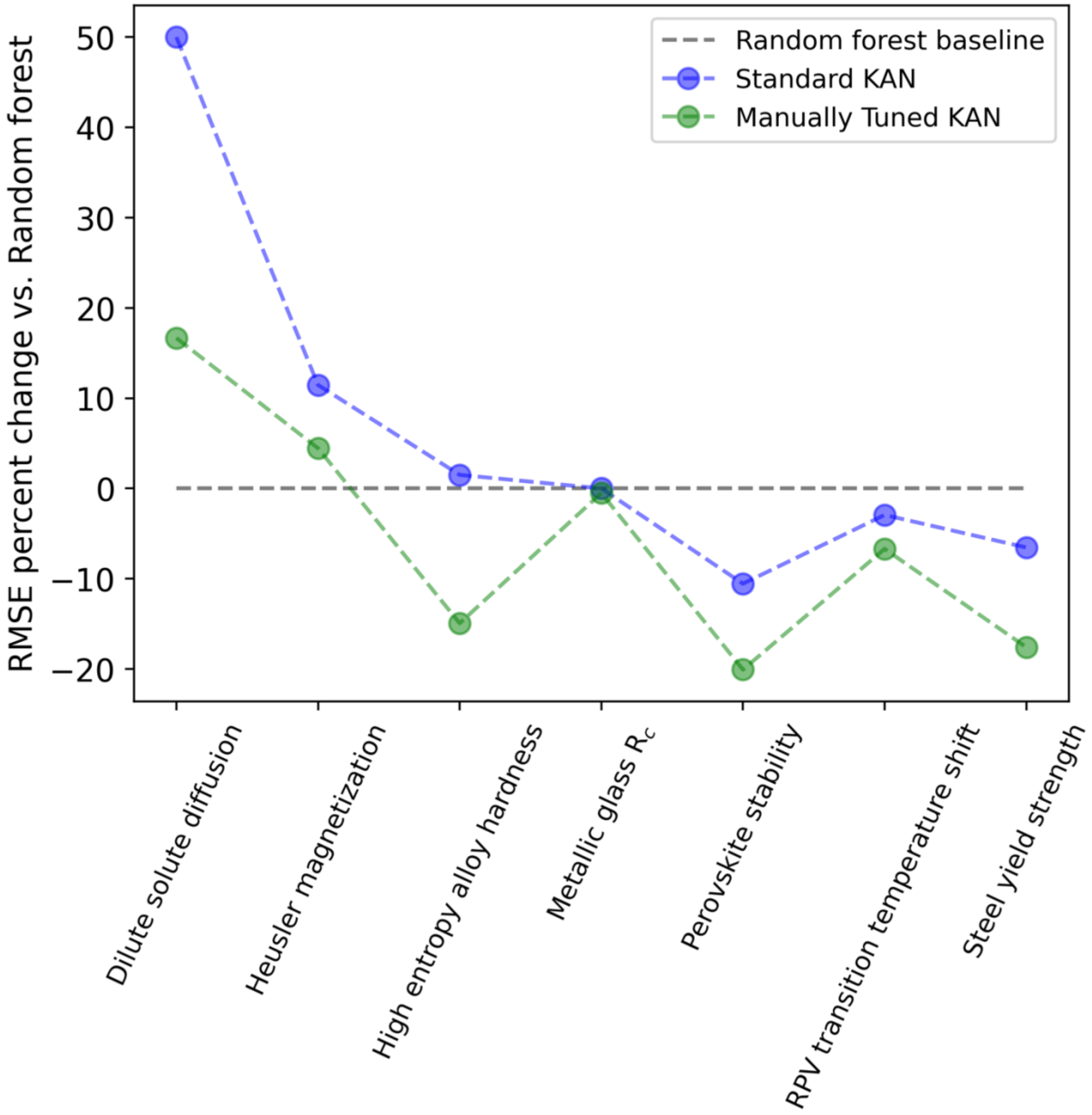

**Figure 2.** Percent change in test data RMSE values for different models on a subset of materials properties. All KAN models were trained for 10 iterations (see values in **Table 3**).

### 3.3. KAN performance by tuning network parameters with Optuna

Next, we explored the impact of a more complete optimization that included activation function hyperparameters, adding additional layers, and tuning the node counts in each layer. To manage this larger optimization task we automated it using the Optuna[55] python package. This analysis was performed to assess the quality of more optimized fits compared to the standard models using the Kolmogorov-Arnold theorem discussed in **Section 3.1** and the manually tuned models discussed in **Section 3.2**. We performed this automated optimization of parameters for 24 of the 33 datasets, while the remaining 9 datasets were omitted due to large computational expense. Overall, we find that through either our automated or manual parameter tuning, that 15 of the 24 properties showed lower errors for the optimized KANs vs. the standard, unoptimized KANs. **Figure 3** contains a summary of ML model fits of these 15 improved KAN fits with the same 2-layer standard KANs and random forest models shown in **Figure 1**. For the optimized KANs, we examined our results from the manual tuning and the automated tuning, and always report the result with the lowest error, which here we just call the "best optimized" KAN. The data points are average RMSE/$\sigma_y$ values from 25 leave out 20% test data splits. The average RMSE/$\sigma_y$ values for random forest, 2-layer standard KAN, and optimized KANs for these 15 datasets were 0.43, 0.53, and 0.44, respectively. We therefore find the average reduction in KAN model error from optimization was about 17%, and, after optimization, average RMSE/$\sigma_y$ over all datasets had essentially the same performance as random forest. Our findings of comparing the performance of standard vs. optimized KANs collectively show that, while KANs in their standard architecture are valuable regression models that can be fit quickly, in practice some significant tuning of parameters and network architecture may be necessary to realize the best performance, where generally the performance improvement from optimization is about 10-20%.

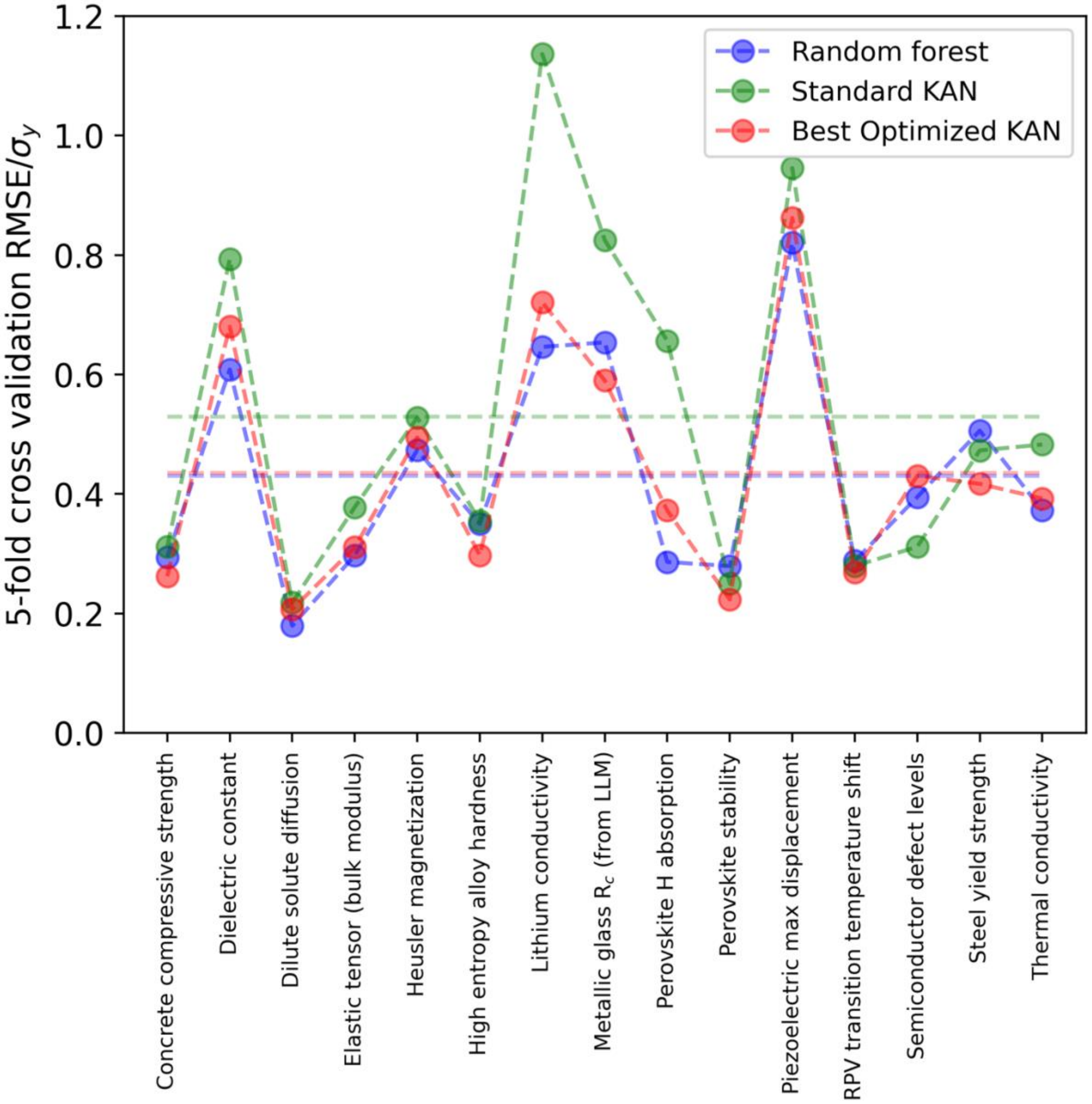

**Figure 3.** Summary of test data RMSE/$\sigma_Y$ for best optimized KAN models (red data), 2-layer standard KAN models (green data) and previously fit random forest models (blue data) for the subset of 15 datasets. The points represent the mean across all test data splits. The horizontal dashed red, green and blue lines are the average RMSE/$\sigma_Y$ values for best optimized KAN, 2-layer standard KAN and RF models, respectively.

Here, we reflect on some practical aspects of fitting KANs. Overall, we find that while standard KANs are capable of being on-par with random forest models, improved KAN models can be obtained through modest tuning of the network node and layer counts. The need for some tuning on the KAN network architecture to realize improved model performance is similar to working with conventional MLPs. However, we have found that the fitting process for KANs is occasionally brittle. More specifically, sometimes numerical issues in fitting lead to very high errors, and cross validation errors with KANs tend to have a wider range than other models like random forest, both of which make KANs more practically difficult to use to obtain stable, robust fits. An advantage of KANs over MLPs is their parameter efficiency and ability to encode specific functional forms for producing simple closed-form models. Overall, our results demonstrate the promise of KANs for materials property prediction, even on small datasets, indicating they should be considered as a standard regression model for any materials ML practitioner. However, when performing materials property prediction on small datasets, we still recommend users to start with workhorse models like random forest or boosted trees, and explore KANs if simplified and highly interpretable models are desirable.

### 3.4. Exploring KANs for predicting RPV embrittlement

To better assess the parameter efficiency and interpretability capabilities of KANs we here consider a single materials property dataset and focus on predicting irradiation-induced embrittlement of reactor pressure vessel (RPV) steels. Prediction of the irradiation-induced embrittlement typically takes the form of predicting a ductile to brittle transition temperature shift (TTS) associated with irradiation based on the steel composition and operational parameters of the reactor. Accurate prediction of TTS is important for understanding embrittlement trends and informing safe and reliable reactor operation.[57–59] The RPV community has traditionally made use of hand-tuned models (often guided by extensive physics-based reasoning) with accessible functional forms of just tens of fitting parameters for predicting embrittlement,[57,59] which we will call "hand-tuned" models here. However, the prediction of RPV steel embrittlement has recently been the focus of numerous ML-based studies,[49,59–63], which offer many advantages in terms of accuracy and ease of development but generally do so at the cost of providing only a black-box model with many fitting parameters (e.g., over $10^6$ parameters in a single neural network model from Jacobs et al.[49]). The capabilities of KANs offer an exciting opportunity to obtain the advantages of machine learning fitting while retaining interpretable, analytical closed-form models. We first create the standard 2-layer KAN which provides a reasonable prediction accuracy, and compare its performance both to larger KAN models and previous work leveraging deep MLP neural networks. We then attempt to create the simplest KAN model that still provides reasonable prediction accuracy, and fit analytical forms to the KAN activation functions to obtain a closed-form model for RPV embrittlement and compare its performance to previously formulated hand-tuned models. [49]

**Figure 4** compares the performance of different KAN models with a previously published deep learning MLP model[49] and an industry-standard hand-tuned TTS model, called E900 (note that the E900 model form was refit to the present data for this analysis).[64] For this analysis, unless otherwise noted, we use the combined TTS database consisting of both test reactor and surveillance data from the work of Jacobs et al.[49] **Figure 4** plots the test data RMSE vs. model parameter count. E900 is the model with fewest parameters (only 26) with an RMSE of 17.5 °C. We were able to make a few very simple KANs (here referred to as "Tiny KAN") which have architectures of 8-1-1, 8-2-1, or 15-1-1, a spline piecewise polynomial of 1 and grid interval of 3, for a total of 27, 48 and 51 parameters, respectively. These Tiny KANs produced RMSEs of 17.2-17.8 °C, essentially identical with the E900 model, and have at most about 2× more parameters. This result is notable for two reasons. First, in terms of neural network complexity, this KAN is very simple and still provides prediction quality rivaling the industry-standard model. Second, the time needed to create and test this KAN was small, on the order of hours (mostly human time trying different fits and debugging code), compared with the many years and deep domain expertise that were used in the development and fine-tuning of the functional forms of the hand-tuned E900 model.

Next, we compare the performance of our different KAN models and previously published deep neural network. Implimenting the standard KAN discussed in **Section 3.1** for this data set leads to a model that contains 2485 parameters and has an RMSE of 13.2 °C. The tuned KAN from **Section 3.2**, referred to here as "Manually tuned KAN", has 10,730 parameters and an RMSE of 12.4 °C. The previously published deep neural network from Jacobs et al.[49] had an RMSE of 12.2 °C, but substantially more parameters than any KAN model investigated here. Jacobs et al.

used an ensemble of deep neural networks for a total of >10M parameters, where a single network contains roughly 1M parameters, as shown in **Figure 4**. It is striking that our tuned KAN and deep neural network attain essentially identical accuracy for a KAN with a factor of 93 × fewer parameters. A modest loss of accuracy of about 5 °C can be achieved with Tiny KAN along with a reduction of more than 20,000× fewer parameters compared to the model from Jacobs et al.[49]

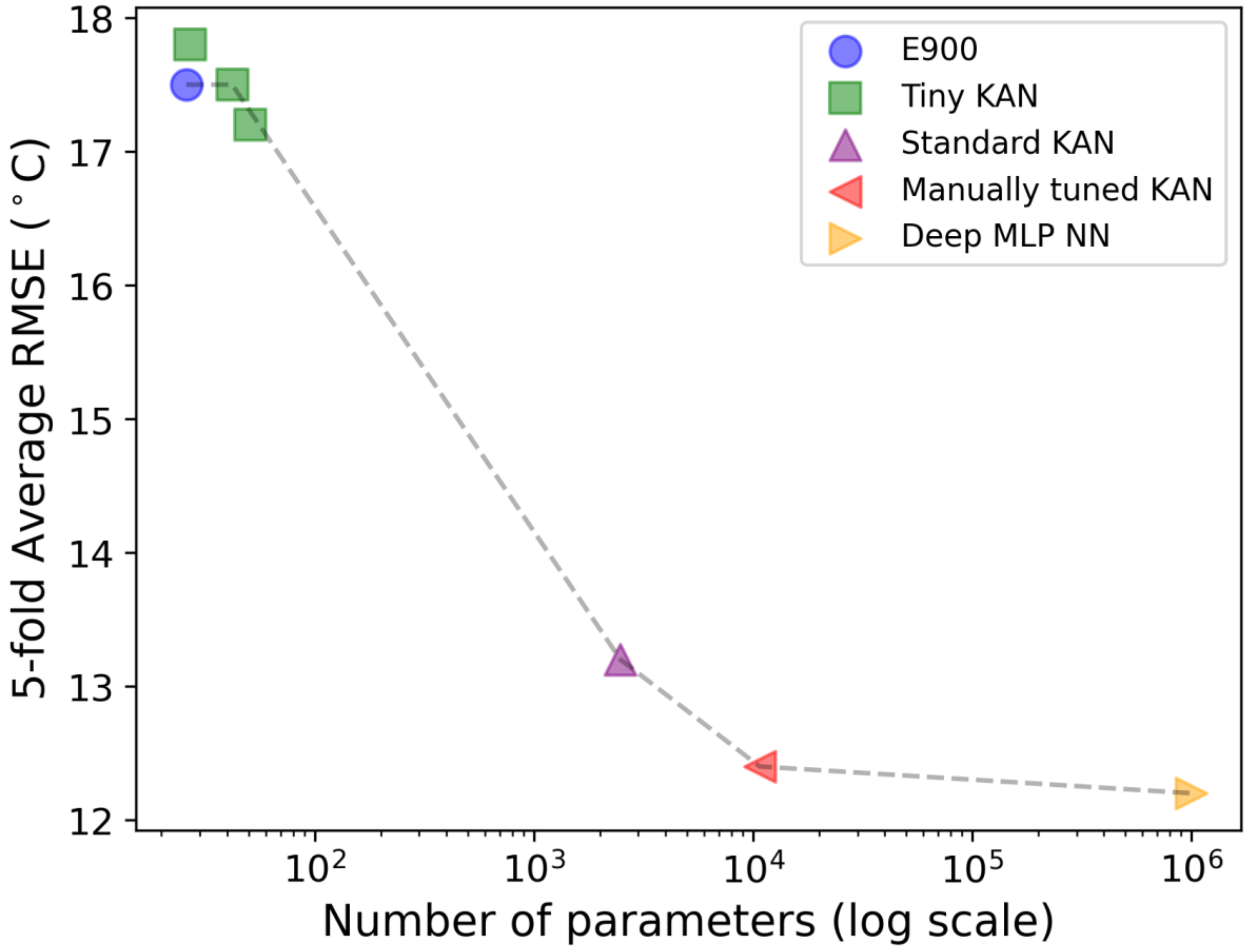

**Figure 4.** Comparison of previous models of RPV embrittlement with various KAN models explored in this work.

Finally, we use our Tiny KAN model to produce a closed-form solution model for RPV embrittlement. The ability of KANs to quickly produce closed-form analytical models holds exciting promise for interpretability and developing models with relatively simple functional dependencies. For this analysis, we restrict ourselves to a subset of the RPV embrittlement data comprising surveillance steels, termed the "Plotter" database in the work of Jacobs et al, as that subset is the data used to develop the E900 model we wish to use for comparison.[49] Previous fits of the E900 model to the Plotter database result in an RMSE of 13.4 °C.[49] Here, we make a KAN with the same 8-2-1 architecture described above, for a total of 48 parameters. This KAN produces a full-fit RMSE of 12.7 °C, slightly better than the E900 fit. By pruning the KAN network to reveal only the most important parameters and fitting the activation functions to closed-form functions, we obtain a closed-form expression containing only 22 adjustable parameters and an RMSE of 14.4 °C. This closed-form model contains only linear, quadratic, exponential, and sine functions, is of comparable in overall RMSE to the E900 model, but was produced in a totally data-centric way, without in-depth domain guidance and knowledge to fine-tune the functional forms. The closed-form expression for predicting embrittlement on the Plotter RPV database is provided in a python notebook available online (see **Data and Code Availability**). We note that our numerical KAN models have been assessed vs. other models like E900 only through comparison of RMSE. Many other metrics, e.g., mean absolute error, extrapolation to high fluence low flux conditions, bias at high-fluence, etc. are important to determining a useful RPV embrittlement model. It is beyond the scope of this work to consider all these aspects. Therefore, these results should be taken only to suggest that KANs could be a useful tool for RPVs embrittlement and other similar modeling, but not to imply that the present KAN models are equal to those already developed for RPV embrittlement.

## 4. Summary and Conclusion:

This study explored the use of KANs in the context of materials property prediction. We broadly find that, even in their standard, unoptimized state, KANs are useful regression models with comparable (although slightly worse on average) performance to other commonly employed methods like random forests. More specifically, we find standard, unoptimized KANs perform on par or better than random forest for about 40% of the materials property datasets examined in this study, with two datasets not converging at all with KANs. If modest optimization is attempted, the prediction errors of KAN models decrease by about 10-20% on average. It is worth noting that even when KANs outperform random forest, the extent of error reduction is mild, where we find an average of less than 10% error reduction vs. random forests. Overall, our findings indicate that KANs in their present form do not stand out as dramatically better than a standard method like random forest, but that they are sometimes better and effective enough on average that they should be considered in the standard repertoire of ML regression models for practitioners interested in creating models of materials property predictions.

In-depth study of KAN predictions of reactor pressure vessel steel embrittlement transition temperature shift (TTS) data led to two important findings. First, an optimized KAN with nearly 100× fewer parameters than a previously fit deep MLP neural network resulted in essentially the same level of accuracy. The resulting KAN model is therefore drastically simpler and faster to train than the MLP neural network. Second, creation of KANs with fewer than 50 parameters resulted in essentially equal prediction performance with established hand-tuned models created using deep domain expertise of the physics of steel embrittlement. These simple KAN models can be parameterized into a closed-form analytical expression, creating a simple, interpretable model with roughly a 10-15% loss in accuracy from the original KAN. Thus, the KAN approach can yield a closed form model similar in overall RMSE accuracy to that obtained with expert hand-tuning in just a few hours using purely data-centric approaches. The KAN model is not quite as good as the hand-tuned one in overall RMSE for TTS, and has not been assessed on other important metrics, but present results suggest that KANs could be a fruitful path for more detailed study of TTS. In general, the embrittlement findings highlight the potential for application of KANs in areas of materials property prediction that value simpler functional forms and involve significant hand-tuning of physics-based analytical models. The results suggest that for such problems KANs might provide a fully data-centric model in a fraction of the time and with less domain expertise than traditional approaches.

**Acknowledgements:**

Support for R. J., L. E. S. and D. M. were provided by the National Science Foundation under NSF Award Number 103173 "Frameworks: Garden: A FAIR Framework for Publishing and Applying AI Models for Translational Research in Science, Engineering, Education, and Industry". Support for R. J. and D. M. for the RPV data analysis was provided by the US Department of Energy (DOE) Nuclear Energy University Program (NEUP) under award number DE-NE0009143.

**Data Availability:**

All datasets used to fit the machine learning models are available on Figshare (https://doi.org/10.6084/m9.figshare.31151638) and are hosted on Foundry-ML (https://foundry-ml.org/#/). Output data for all KAN fits and an example python notebook to run the KAN model on the RPV embrittlement dataset, and the resulting closed-form analytical formula, is available on Figshare (https://doi.org/10.6084/m9.figshare.31151638). The MAST-ML code is available on Github (https://github.com/uw-cmg/MAST-ML) and installable via pip.

**Conflicts of Interest**

The authors have no conflicts of interest to declare.